\documentclass[singlecolumn,aps,pra,showpacs,superscriptaddress,nofootinbib]{revtex4-2}
\usepackage{graphicx} % Required for inserting images
\usepackage{color}
\usepackage[utf8]{inputenc}
\usepackage{amsmath,amsfonts,amssymb}
\usepackage[colorlinks]{hyperref}
\usepackage{tikz,pgfplots}\pgfplotsset{compat=1.18}

\begin{document}

\title{Tighter Asymptotic Key Rates for Intensity-Correlated Decoy-State QKD via Nonlinear Programming}
\author{Matej Pivoluska}
\email{mp@qtl.at}
\affiliation{%
 qtlabs GmbH - Quantum Technology Laboratories GmbH, Clemens-Holzmeister-Straße 6/6 Etage 6, 1100 Wien, Austria
}%
\author{Mateus Araújo}
\affiliation{Departamento de Física Teórica, Atómica y Óptica, Laboratory for Disruptive Interdisciplinary Science (LaDIS), Universidad de Valladolid, 47011 Valladolid, Spain}
\date{\today}
\begin{abstract}
Decoy-state QKD with phase-randomized weak coherent pulses is typically analyzed assuming independent, precisely prepared intensities. Real sources, however, can exhibit correlated intensity drift across rounds, potentially leaking intensity information and breaking the standard decoy-state reduction to linear programs. Cauchy--Schwarz (CS) constraints can restore security by coupling $n$-photon yields across intensities, but they introduce nonlinear square-root constraints that are commonly handled via outer linearisation around channel-model-based reference points. We propose a reproducible alternative: first solve the full CS-constrained parameter-estimation problems using the interior-point nonlinear solver IPOPT, then use the resulting candidate solution as the linearisation point for the outer optimisation that certifies a valid lower bound on the asymptotic key rate. Simulations for both coarse-grained model-independent correlations and fine-grained truncated-Gaussian models show consistently tighter key-rate bounds than canonical reference points, and in some cases allow certifying optimality when both optimisation stages coincide.
\end{abstract}

\maketitle

\section{Introduction}
Quantum key distribution (QKD) is a cryptographic protocol designed to create secret and random shared keys between users. 
Such keys can later be used to build other cryptographic primitives, such as a secret \cite{Vernam1926} and authenticated \cite{Wegman1981} channel. 
The fact that QKD {}{typically} does not rely on any computational assumptions uniquely positions it in the current cryptographic landscape because, unlike other conventional key distribution schemes, it is naturally resistant to both quantum computers and any future advances in classical algorithms \cite{Renner2023a}.

On the flip side, most of the practical QKD protocols require extensive modeling of the devices used in their implementation, and the current device models do not describe the practical devices well enough.
More concretely, most of the current security proofs are constructed around the assumption that the QKD source can prepare the states prescribed by the protocol to infinite precision and that the measurement device exactly implements the prescribed POVMs.  
In practice, this is far from the truth, and hence a significant effort was invested into loosening these assumptions and allowing more realistic descriptions for both QKD sources \cite{Wang2019b,Mizutani2021, Nahar2023a,CurrasLorenzo2023,Pereira2025,Nagamatsu2016,Pereira2020,Zapatero2021,CurrasLorenzo2023a,Pereira2024a,Marwah2024,Li2025c,Mizutani2019a,Sixto2022a,Li2025} and QKD measurement devices \cite{Burenkov2010, Kamin2024,Tupkary2024,Marcomini2025a,Grasselli2025,Nahar2025,Wang2025f}.

In this paper, we contribute to this effort and discuss a particular source imperfection in the context of a decoy state QKD \cite{Lo2005,Ma2005,Wang2005}.
In decoy state QKD, due to the lack of single photon sources, the information is encoded in phase randomized weak coherent pulses (PRWCP) of varying intensities.

The usual assumption is that the PRWCPs can be prepared with exact intensities independently in each round of the protocol.
However, in practical devices, this is not the case, and the intensity of the pulses can be correlated non-trivially over multiple protocol rounds \cite{Pereira2020}.
This imperfection has previously been addressed by multiple studies in the asymptotic limit for various correlation models \cite{Zapatero2021,Sixto2022a,Li2025}.

The core of the decoy state QKD is estimating the relevant parameters conditioned on a single photon emission (which are not directly accessible through measurement) using the measured information conditioned on intensities (which are directly accessible).
The standard decoy state method is based on the fundamental assumption that the probability of detection of a signal with $n$ photons (often called ``yield'') depends only on the photon number $n$ and not the intensity setting used.
Under this assumption the \emph{decoy state method} can be cast as a series of linear programs (LPs) and the relevant parameters conditioned on single photons being sent can be found by solving the \cite{Rice2009}.
Unfortunately, this assumption is violated in the QKD sources with intensity correlations, because information about the intensity setting in each round is potentially leaked via intensities of the subsequent pulses, which allows the potential adversary to modify their eavesdropping strategy, which can lead to the dependence of the yields on the intensities used as well as the photon number.

In \cite{Zapatero2021} a modification of the decoy state LP was proposed to address this issue. Additional constraints -- called the Cauchy-Schwartz (CS) constraints -- restricting how much the intensity dependent $n$-photon yields can differ from each other was added to the decoy state LPs. 
One technical aspect of this addition is that the resulting optimisation problem is not an LP anymore, because CS constraints are not linear functions of the optimisation variables. Fortunately, the CS-constraints are at least convex functions of the optimisation variables and hence to solve this optimisation the existing literature \cite{Zapatero2021,Sixto2022a,Li2025,Trefilov2025} uses an outer optimisation over this convex set defined by replacing the CS constraints by \emph{linearized CS constraints}, which are simply derivatives of the original CS constraints in chosen \emph{reference point}. 
Such outer optimisation is a LP and hence can be efficiently solved.
Because the outer optimisation simply enlarges the search space it always leads to valid (but possibly suboptimal) key rates for any choice of the reference points. 
At the same time canonical reference points were proposed -- chosen according to a noise model theoretically describing the quantum channel for which the achievable asymptotic key rate is being estimated. 

In this paper we propose an alternative solution to the choice of the reference points for calculation of the asymptotic key rate for QKD implementations with intensity correlations: we solve the original problems with non-linearized CS-constraints to obtain a candidate solutions $\vec{\mathcal{S}}$ with the interior-point smooth non-linear solver IPOPT \cite{waechter2006}. While these candidate solutions are not guaranteed to be optimal, they often are. In order to obtain true lower bounds for the key rate we linearize the problem using these reference points, as before.

This new approach has two advantages. First, optimality can be certified in certain cases. 
If both optimisation rounds return the same result, it means the problem with non-linearized CS constraints was solved correctly and the resulting choice of the linearization points as well as the key rate is optimal.
Second, even in cases where the first optimization does not successfully find the optimal solution, the method produces linearization points that outperform the canonical linearization points chosen according to the channel model {}{(see Fig. \ref{fig:canonical vs IPOPT} for comparison to results reported in \cite{Zapatero2021})}. 
This is especially true in cases where the channel model does not faithfully reflect the actual quantum channel, such as in some of the current studies using this technique to explore properties of real experimental devices \cite{Trefilov2025}
{}{(see Fig. \ref{fig:Trefilov} for comparison)}. 

The remainder of the paper is organized as follows.
In section \ref{sec:TechnicalIntro} we review the decoy state methods for PRWCP sources with correlated intensities and discuss different intensity correlation models introduced in the literature.
In section \ref{sec:simulations} we apply our new method to re-derive key rates from data reported in recent publications \cite{Zapatero2021,Trefilov2025} using our new method. 
Thus, we demonstrate that our method can both achieve substantially better key rates and certify the achieved key rates are optimal in some cases.

\section{Key-rate optimisation with intensity-correlated sources}\label{sec:TechnicalIntro}

In this section we review the methods to calculate the asymptotic key rate for decoy state QKD implemented with correlated PRWCP sources and formulate the associated parameter-estimation problems as non-linear optimization tasks. 

%These optimisation problems are then solved with IPOPT, which allows us to keep the full nonlinear Cauchy--Schwarz (CS) constraints that couple different intensity settings, instead of resorting to outer linearisation as in the original treatment by Zapatero \emph{et al.} 
This section also introduces the channel model used to generate simulated data in some of the existing literature and reference points to perform the outer optimization. Finally, we outline different ways of modeling intensity correlations: a coarse-grained, model-independent description \cite{Zapatero2021} and a fine-grained description with a truncated-Gaussian correlation model inspired by experimental characterisation of real sources \cite{Sixto2022a}. 

\subsection{Protocol description}

For concreteness, we consider a standard polarization-encoding, decoy-state BB84 protocol. In each round $k$, the sender (Alice) chooses:
\begin{itemize}
    \item a basis $x_k \in M = \{X,Z\}$ with probability $q_{x_k}$,
    \item a raw key bit $r_k \in \mathbb{Z}_2 = \{0,1\}$, uniformly at random,
    \item an intensity setting $a_k \in A = \{\mu,\nu,\omega\}$ with probability $p_{a_k}$.
\end{itemize}
The probabilities $q_{x_k}$ and $p_{a_k}$ depend only on the chosen basis and intensity, respectively, and not on the round index $k$. Alice encodes the BB84 state defined by $(x_k,r_k)$ into a phase-randomised weak coherent pulse (PRWCP) with nominal mean photon number $a_k$, and sends it through the quantum channel to the receiver (Bob).

Because of imperfections in the source, the actual mean photon number $\alpha_k$ of the emitted pulse need not coincide with the chosen intensity setting $a_k$. These deviations are correlated across different rounds, so that in general the distribution of $\alpha_k$ may depend on several previous intensity choices. Throughout this section we assume:
\begin{itemize}
    \item perfect phase randomisation,
    \item no state-preparation flaws beyond the intensity drift,
    \item no side-channels.
\end{itemize}

On Bob’s side, in each round $k$ he chooses a basis $y_k \in M$ with probability $q_{y_k}$, again independent of the round index, and performs a POVM on the incoming pulse $B_k$,
\[
\{M^{y_k,s_k}_{B_k}\}_{s_k\in\{0,1,f\}},
\]
where $s_k \in \{0,1\}$ denotes a conclusive outcome and $s_k=f$ denotes a no-click event. We assume basis-independent detection efficiency, so that
\[
M^{Z,f}_{B_k} = M^{X,f}_{B_k} \equiv M^{f}_{B_k}.
\]

This process produces measurement data, which is then sifted (only detected signals in matched basis are kept) and the decoy state method is applied to estimate the parameters necessary to evaluate the asymptotic key rate (see below).

\subsection{Modelling intensity correlations}

We now summarise the assumptions on the statistics of the actual intensity $\alpha_k$ and its correlations across rounds as introduced in \cite{Zapatero2021}.
The {}{Assumptions in subsequent works \cite{Sixto2022a,Li2025,Trefilov2025} are practically the same, however they differ in the detail of modeling the correlation function $g_{\vec{a}_k}(\alpha_k)$ (see section \ref{sec:truncatedGaussianModel} for a description of the truncated Gaussian correlation model used in \cite{Sixto2022a,Trefilov2025}), which allows to tighten the calculation of the photon number statistics in Eq.(3). Our method is applicable to all these works.}

\paragraph{Assumption 1 (Poisson photon-number statistics).}
Conditioned on the actual intensity $\alpha_k$ in round $k$, the photon-number distribution is Poissonian:
\begin{equation}
p(n_k|\alpha_k) = \frac{e^{-\alpha_k}\alpha_k^{n_k}}{n_k!}, \qquad \forall n_k\in\mathbb{N}. \label{eq:assumption1}
\end{equation}

\paragraph{Assumption 2 (bounded relative deviation).}
For each possible record of previous intensity settings $\vec{a}_k = (a_k, a_{k-1},\dots)$, the actual intensity $\alpha_k$ is supported only on an interval around the nominal setting,
\begin{equation}
\left\vert 1-\frac{\alpha_k}{a_k}\right\vert\leq\delta_\text{max}, \label{eq:assumption2-1}
\end{equation}
i.e.\ the conditional density $g_{\vec{a}_k}(\alpha_k)$ is nonzero only for
\[
\alpha_k \in [a_k^{-},a_k^{+}], \qquad a_k^{\pm}=a_k(1\pm\delta_\text{max}).
\]
Thus $\delta_\text{max}$ bounds the maximal relative deviation between the nominal intensity $a_k$ and the actual intensity $\alpha_k$.

Combining Assumptions 1 and 2, the photon-number statistics in round $k$, conditioned on the record $\vec{a}_k$, are
\begin{equation}
\left.p_{n_k}\right|_{\vec{a}_k}
=\int_{a_k^{-}}^{a_k^{+}} g_{\vec{a}_k}\left(\alpha_k\right) 
\frac{e^{-\alpha_k} \alpha_k^{n_k}}{n_{k}!} d \alpha_k,
\qquad \forall n_k\in\mathbb{N}. \label{eq:assumption2-2}
\end{equation}

\paragraph{Assumption 3 (finite correlation range).}
We assume that the intensity correlations have a finite memory length $\xi$. That is, $g_{\vec{a}_k}(\alpha_k)$ depends only on the last $\xi$ intensity settings,
and is independent of $a_j$ for which $k-j>\xi$.

Beyond these three assumptions, the function $g_{\vec{a}_k}(\alpha_k)$ remains completely unknown, making the analysis model-independent. Instead of specifying $g_{\vec{a}_k}$, we work with bounds on the photon-number probabilities implied by Assumptions~1 and~2. 

\subsection{Photon-number distribution with correlated intensities}

In the presence of correlations, the photon-number distribution in round $k$ depends not only on the current intensity setting $a_k=a$, but also on the sequence of previous settings $\vec{a}_{k-1}=(a_{k-1},\dots,a_1)$:
\begin{equation}
p^{(k)}(n|a) 
= \sum_{\vec{a}_{k-1}}
\bigl( p_{a_{k-1}}p_{a_{k-2}}\cdots p_{a_1} \bigr)\,
p_{n|a,\vec{a}_{k-1}}. \label{eq:assumption3-1}
\end{equation}
Here $p_{n|a,\vec{a}_{k-1}}$ is the photon-number distribution in round $k$ given the current setting $a_k=a$ and the past record $\vec{a}_{k-1}$, which is of the form presented in Eq.~\eqref{eq:assumption1}.

Using Eq.~\eqref{eq:assumption2-1} and the monotonicity properties of $e^{-x}x^n$ on $(0,1)$, one can derive \emph{record-independent} bounds on $p^{(k)}(n|a)$:
\begin{align}
p^{(k)}(0|a) &= p(0|a) \in \left[e^{-a^+}, e^{-a^-}\right], \label{eq:probabilityBounds1}\\\nonumber
p^{(k)}(n|a) &= p(n|a) \in \left[\frac{e^{-a^-}(a^-)^n}{n!},\, \frac{e^{-a^+}(a^+)^n}{n!}\right], 
\qquad n\geq 1,
\end{align}
where $a^\pm = a(1\pm\delta_\text{max})$.
In the following we denote these bounds compactly as
\[
p(n|a)\in [p_L(n|a),p_U(n|a)].
\]

\subsection{Key-rate formula}

The asymptotic secret-key rate per signal in the $Z$-basis is given by \cite{Zapatero2021}:
\begin{equation}
K_{\infty}
={Z}_{1, \mu}^{\mathrm{L}}
\left[1-h\left(\frac{{E}_{1, \mu}^{\mathrm{U}}}{{X}_{1, \mu}^{\mathrm{L}}}\right)\right]
-f_{\mathrm{EC}} {Z}_{\mu} h\left(E_{\mathrm{tol}}\right), \label{eq:keyRateFormula}
\end{equation}
where:
\begin{itemize}
    \item $h(\cdot)$ is the binary entropy function.
    \item $f_{\mathrm{EC}}$ is the error-correction efficiency.
    \item $E_{\mathrm{tol}}$ is the tolerated bit-error rate in the raw $Z$-basis key. 
    \item $Z_{\mu}$ is the observed fraction of detected signals (``clicks'') when both Alice and Bob use the $Z$ basis and Alice chooses intensity $\mu$.
    \item $Z_{1,\mu}^{\mathrm{L}}$ is a lower bound on the fraction of detected $Z$-basis signals for which Alice used intensity $\mu$ and a single photon was emitted.
    \item $X_{1,\mu}^{\mathrm{L}}$ is the analogous lower bound for the $X$ basis.
    \item $E_{1,\mu}^{\mathrm{U}}$ is an upper bound on the error rate in the $X$ basis when the intensity $\mu$ is used and a single photon is emitted.
\end{itemize}
The quantities $Z_\mu$ and the corresponding $X_\mu$ are directly observable in experiment or computable from a channel model. 
In contrast, $Z_{1,\mu}$, $X_{1,\mu}$, and $E_{1,\mu}$ depend on the single-photon statistics and are not directly accessible. They must be bounded via the decoy state methods using constrained optimisation.

\subsection{Yields, error rates, and observable quantities}

For each photon number $n\in\mathbb{N}$, intensity setting $c\in A$, and bit value $r\in\{0,1\}$ we define the basis-dependent yields
\begin{align}
Y^Z_{n,c} &= p(\text{click}\mid n, c, Z, Z), \\
Y^X_{n,c} &= p(\text{click}\mid n, c, X, X),
\end{align}
and the error probability in the $X$ basis when bit $r$ is encoded:
\begin{equation}
H^X_{n,c,r} = p(\text{err}\mid n,c,X,X,r).
\end{equation}
Here we use shorthand notation such as
\[
p(\text{click}\mid n,c,Z,Z) 
= p(s_k\neq f\mid n_k=n,a_k=c,x_k=Z,y_k=Z).
\]
We also define the bit-averaged error probability in $X$:
\begin{equation}
H^X_{n,c} = \frac{H^X_{n,c,0} + H^X_{n,c,1}}{2}.
\end{equation}
All these quantities are \emph{not} directly observable, because the photon number $n$ is not measured in the protocol.

The single-photon contributions entering the key-rate formula can be written as
\begin{align}
Z_{1,\mu} &= q_Z^2\, p_\mu\, p(1|\mu)\, Y^Z_{1,\mu}, \\
X_{1,\mu} &= q_X^2\, p_\mu\, p(1|\mu)\, Y^X_{1,\mu}, \\
E_{1,\mu} &= q_X^2\, p_\mu\, p(1|\mu)\, H^X_{1,\mu}.
\end{align}
We therefore need lower bounds on $Y^Z_{1,\mu}$ and $Y^X_{1,\mu}$ and an upper bound on $H^X_{1,\mu}$, consistent with the observed click and error statistics.

For each intensity $a\in A$ we define the observable probabilities
\begin{align}
Z_a &= p(a,Z,Z,\text{click})
= q_Z^2 p_a \sum_{n=0}^\infty p(n|a)\,Y^Z_{n,a}, \\
X_a &= p(a,X,X,\text{click})
= q_X^2 p_a \sum_{n=0}^\infty p(n|a)\,Y^X_{n,a}, \\
E_a &= p(a,X,X,\text{err})
= q_X^2 p_a \sum_{n=0}^\infty p(n|a)\,H^X_{n,a}.
\end{align}

\subsection{Standard decoy-state method without intensity correlations}

In the usual decoy-state scenario without intensity correlations, the eavesdropper has no information about the intensity choice in each round; the yields and error rates then depend only on $n$, not on the specific intensity $a$. Therefore one has
\[
\forall a,b\in A:
\quad
Y^Z_{n,a} = Y^Z_{n,b}, \quad
Y^X_{n,a} = Y^X_{n,b}, \quad
H^X_{n,a} = H^X_{n,b}.
\]
In this case, to bound $Y^Z_1$ one may solve a simple optimization problem expressed as
\begin{equation}
\begin{aligned}
& \min\; Y^Z_1 \\
& \text{subject to}\\
& Z_a 
= q_Z^2 p_a \sum_{n=0}^\infty p(n|a)\,Y^Z_n,
\quad \forall a\in A.
\end{aligned}
\end{equation}
To make the problem finite-dimensional, one introduces a photon-number cut-off $N_\text{cut}$ and bounds the contribution from $n>N_\text{cut}$:
\begin{equation}
\begin{aligned}
& \min\; Y^Z_1 \\
& \text{subject to}\\
& Z_a \geq q_Z^2 p_a \sum_{n=0}^{N_\text{cut}} p(n|a)\,Y^Z_n,
\quad \forall a\in A,\\
& Z_a \leq q_Z^2 p_a \left[
\sum_{n=0}^{N_\text{cut}} p(n|a)\,Y^Z_n
+ 1-\sum_{n=0}^{N_\text{cut}}p(n|a)
\right],\quad \forall a\in A.
\end{aligned}
\end{equation}
Analogous linear programs can be written for $Y^X_1$ and $H^X_1$, yielding bounds that can be plugged into the key-rate formula Eq.~\eqref{eq:keyRateFormula}.

\subsection{Breakdown of the standard method with intensity correlations}

In the presence of intensity correlations the yields and error rates acquire an explicit dependence on the intensity setting $a$.
Trying to generalise the previous linear program to this case leads to
\begin{equation}
\begin{aligned}
& \min\; Y^Z_{1,\mu} \\
& \text{subject to}\\
& Z_a \geq q_Z^2 p_a \sum_{n=0}^{N_\text{cut}} p(n|a)\,Y^Z_{n,a},
\quad \forall a\in A,\\
& Z_a \leq q_Z^2 p_a \left[
\sum_{n=0}^{N_\text{cut}} p(n|a)\,Y^Z_{n,a}
+ 1-\sum_{n=0}^{N_\text{cut}}p(n|a)
\right], \quad \forall a\in A.
\end{aligned}
\end{equation}
However, it is easy to see that the constraints for $a\neq\mu$ become effectively irrelevant for bounding $Y^Z_{1,\mu}$: the different intensities decouple, and we can severely underestimate $Y^Z_{1,\mu}$ while still satisfying all constraints. Additional constraints are therefore needed to link the photon-number statistics associated with different intensities.

Zapatero \emph{et al.} \cite{Zapatero2021} introduced a family of Cauchy--Schwarz (CS) constraints that correlate the yields and error rates across intensities. 
These constraints are nonlinear and are the main source of complexity in the resulting optimisation problems. 
As mentioned in the introduction, in the original work they are treated by an outer linearisation; in our approach we keep them in their nonlinear form and solve the resulting nonlinear programs results of which are then used as reference point for outer linearization.

\subsection{Coarse-grained model-independent CS constraintsformulation}\label{subsec:coarseGrainedFormulation}

The CS constraints relate $Y_{n,a}$ and $Y_{n,b}$ for any two intensities $a,b\in A$ and similarly for $H_{n,a}$ and $H_{n,b}$. For each photon number $n$ they read
\begin{equation}\label{eq:CSconstraints}
\begin{aligned}
& G_{-}\left(Y_{n, a}, \tau_{a b, n}^{\xi}\right) \leq Y_{n, b} \leq G_{+}\left(Y_{n, a}, \tau_{a b, n}^{\xi}\right), \\
& G_{-}\left(H_{n, a}, \tau_{a b, n}^{\xi}\right) \leq H_{n, b} \leq G_{+}\left(H_{n, a}, \tau_{a b, n}^{\xi}\right),
\end{aligned}
\end{equation}
where
\begin{equation}\label{eq:split}
\begin{aligned}
G_{-}(y, z)&=\begin{cases}
g_{-}(y, z) & \text{if } y>1-z, \\
0 & \text{otherwise},
\end{cases}
\\
G_{+}(y, z)&=\begin{cases}
g_{+}(y, z) & \text{if } y<z, \\
1 & \text{otherwise},
\end{cases}
\end{aligned}
\end{equation}
with
\begin{equation}\label{eq:sqrt}
g_{ \pm}(y, z)
=y+(1-z)(1-2 y) \pm 2 \sqrt{z(1-z) y(1-y)}. 
\end{equation}
The parameter $\xi$ is the correlation range, and
\begin{equation}\label{eq:coarseGrainedTau}
\begin{aligned}
\tau_{a b, n}^{\xi}= 
\begin{cases}
e^{a^{-}+b^{-}-\left(a^{+}+b^{+}\right)}
\left[1-\sum_{c \in A} p_c\left(e^{-c^{-}}-e^{-c^{+}}\right)\right]^{2 \xi} & \text{if } n=0, \\
e^{a^{+}+b^{+}-\left(a^{-}+b^{-}\right)}
\left(\dfrac{a^{-} b^{-}}{a^{+} b^{+}}\right)^n
\left[1-\sum_{c \in A} p_c\left(e^{-c^{-}}-e^{-c^{+}}\right)\right]^{2 \xi} & \text{if } n \geq 1 .
\end{cases}
\end{aligned}
\end{equation}

Using the bounds $p(n|a)\in[p_L(n|a),p_U(n|a)]$ and the CS constraints~\eqref{eq:CSconstraints}, the three main optimisation problems needed for the key-rate formula~\eqref{eq:keyRateFormula} can be written as follows.

\paragraph{Problem (P1): lower bound on $Z_{1,\mu}$.}
\begin{equation}
\begin{aligned}
&\min\; Y^Z_{1,\mu} \\
&\text{subject to:}\\[1mm]
&\frac{Z_a}{q_{\mathrm{Z}}^2 p_a}\geq 
e^{-a^+}Y^Z_{0,a}
 + \sum_{n=1}^{N_\text{cut}} \frac{e^{-a^-} (a^{-})^n}{n!} Y^Z_{n,a},
\quad \forall a\in A,\\[1mm]
&\frac{Z_a}{q_{\mathrm{Z}}^2 p_a }\leq 
e^{-a^-}Y^Z_{0,a}
+\sum_{n=1}^{N_\text{cut}} \frac{e^{-a^+} (a^{+})^n}{n!} Y^Z_{n,a}
+\left(1-e^{-a^+}-\sum_{n=1}^{N_\text{cut}}\frac{e^{-a^+} (a^{+})^n}{n!} \right),
\quad \forall a\in A,\\[1mm]
& G_{-}\left(Y^Z_{n, a}, \tau_{a b, n}^{\xi}\right) 
\leq Y^Z_{n, b} 
\leq G_{+}\left(Y^Z_{n, a}, \tau_{a b, n}^{\xi}\right), 
\quad\forall a,b \in A,\ a\neq b,\ n=0,\dots,N_\text{cut}, \\[1mm]
&0\leq Y^Z_{n,a}\leq 1, \quad \forall a\in A,\ n = 0,\dots,N_\text{cut}.
\end{aligned}
\end{equation}

\paragraph{Problem (P2): lower bound on $X_{1,\mu}$.}
\begin{equation}
\begin{aligned}
&\min\; Y^X_{1,\mu} \\
&\text{subject to:}\\[1mm]
&\frac{X_a}{q_{\mathrm{X}}^2 p_a}\geq 
e^{-a^+}Y^X_{0,a}
 + \sum_{n=1}^{N_\text{cut}} \frac{e^{-a^-} (a^{-})^n}{n!} Y^X_{n,a},
\quad \forall a\in A,\\[1mm]
&\frac{X_a}{q_{\mathrm{X}}^2 p_a }\leq 
e^{-a^-}Y^X_{0,a}
+\sum_{n=1}^{N_\text{cut}} \frac{e^{-a^+} (a^{+})^n}{n!} Y^X_{n,a}
+\left(1-e^{-a^+}-\sum_{n=1}^{N_\text{cut}}\frac{e^{-a^+} (a^{+})^n}{n!} \right),
\quad \forall a\in A,\\[1mm]
& G_{-}\left(Y^X_{n, a}, \tau_{a b, n}^{\xi}\right) 
\leq Y^X_{n, b} 
\leq G_{+}\left(Y^X_{n, a}, \tau_{a b, n}^{\xi}\right), 
\quad\forall a,b \in A,\ a\neq b,\ n=0,\dots,N_\text{cut}, \\[1mm]
&0\leq Y^X_{n,a}\leq 1, \quad \forall a\in A,\ n = 0,\dots,N_\text{cut}.
\end{aligned}
\end{equation}

\paragraph{Problem (P3): upper bound on $E_{1,\mu}$.}
\begin{equation}
\begin{aligned}
&\max\; H^X_{1,\mu} \\
&\text{subject to:}\\[1mm]
&\frac{E_a}{q_{\mathrm{X}}^2 p_a}\geq 
e^{-a^+}H^X_{0,a}
 + \sum_{n=1}^{N_\text{cut}} \frac{e^{-a^-} (a^{-})^n}{n!} H^X_{n,a},
\quad \forall a\in A,\\[1mm]
&\frac{E_a}{q_{\mathrm{X}}^2 p_a }\leq 
e^{-a^-}H^X_{0,a}
+\sum_{n=1}^{N_\text{cut}} \frac{e^{-a^+} (a^{+})^n}{n!} H^X_{n,a}
+\left(1-e^{-a^+}-\sum_{n=1}^{N_\text{cut}}\frac{e^{-a^+} (a^{+})^n}{n!} \right),
\quad \forall a\in A,\\[1mm]
& G_{-}\left(H^X_{n, a}, \tau_{a b, n}^{\xi}\right) 
\leq H^X_{n, b} 
\leq G_{+}\left(H^X_{n, a}, \tau_{a b, n}^{\xi}\right), 
\quad\forall a,b \in A,\ a\neq b,\ n=0,\dots,N_\text{cut}, \\[1mm]
&0\leq H^X_{n,a}\leq 1, \quad \forall a\in A,\ n = 0,\dots,N_\text{cut}.
\end{aligned}
\end{equation}

The only nonlinearities in (P1)--(P3) come from the CS constraints~\eqref{eq:CSconstraints}, which contain square-root terms. All other constraints are linear in the optimisation variables. 

\subsection{Outer linearisation of CS constraints}

For comparison with the approach of Zapatero \emph{et al.} \cite{Zapatero2021}, we briefly recall how the CS constraints are treated by outer linearisation. In their scheme, the nonlinear inequalities~\eqref{eq:CSconstraints} are replaced by first-order Taylor expansions around some reference yields $\tilde{Y}_{n,a}\in(0,1)$ (and similarly for $\tilde{H}_{n,a}$). The nonlinear bounds
\[
G_{-}\left(Y_{n, a}, \tau_{a b, n}^{\xi}\right) \leq Y_{n, b} \leq 
G_{+}\left(Y_{n, a}, \tau_{a b, n}^{\xi}\right)
\]
are replaced by
\begin{equation}
\begin{aligned}
& G_{-}\left(\tilde{Y}_{n, a}, \tau_{a b, n}^{\xi}\right)
+G^{\prime}_-\left(\tilde{Y}_{n, a}, \tau_{a b, n}^{\xi}\right)\left(Y_{n, a}-\tilde{Y}_{n, a}\right) \leq Y_{n, b} \\
&\leq 
G_{+}\left(\tilde{Y}_{n, a}, \tau_{a b, n}^{\xi}\right)
+G^{\prime}_+\left(\tilde{Y}_{n, a}, \tau_{a b, n}^{\xi}\right)\left(Y_{n, a}-\tilde{Y}_{n, a}\right),
\end{aligned}
\end{equation}
where
\begin{equation}
\begin{aligned}
G^{\prime}_-(y, z)&=\begin{cases}
g^{\prime}_{-}(y, z) & \text{if } y>1-z, \\
0 & \text{otherwise},
\end{cases}
\\
G^{\prime}_+(y, z)&=\begin{cases}
g^{\prime}_+(y,z) &\text{if } y<z, \\
0 & \text{otherwise},
\end{cases}
\end{aligned}
\end{equation}
and
\begin{equation}
g^{\prime}_{ \pm}(y, z)
= -1+2z \pm (1-2y)\sqrt{\frac{z(1-z)}{y(1-y)}}.
\end{equation}
This linearisation is applied to $Y^Z$, $Y^X$ and $H^X$ alike and yields purely linear programs that can be solved by standard LP solvers. However, the quality of the bounds depends on the choice of the reference points $\tilde{Y}_{n,a}$ and $\tilde{H}_{n,a}$ and on how tight the linearisation is.

\subsection{Channel model and reference points}

To generate meaningful simulation data, we specify a simple but realistic channel and detection model (also used in \cite{Zapatero2021,Sixto2022a}. The basic parameters are:
\begin{itemize}
    \item $\eta_\text{ch}$: probability that a photon is transmitted through the channel.
    \item $\delta_A$: misalignment angle in Alice’s preparation or in the channel.
    \item $p_d$: dark-count probability of each detector.
    \item $\eta_\text{det}$: efficiency of each detector.
\end{itemize}
We define the overall transmittance
\[
\eta = \eta_\text{ch}\eta_\text{det}.
\]

Neglecting dark counts for the moment, and assuming that a photon in a polarization that ideally triggers detector $D_1$ is sent, the probabilities for the four detection patterns
\[
00,01,10,11
\]
(no clicks, click in $D_1$, click in $D_2$, double click) when $n$ photons are impinging on Bob are
\begin{align}
p_{00} &= (1-\eta)^n, \\
p_{10} &= \bigl(\eta \cos^2\delta_A + 1-\eta\bigr)^n - (1-\eta)^n, \\
p_{01} &= \bigl(\eta \sin^2\delta_A + 1-\eta\bigr)^n - (1-\eta)^n, \\
p_{11} &= 1 - p_{00} - p_{01} - p_{10}.
\end{align}
Dark counts are independent of the incoming photons and are incorporated afterwards.
Let
$$
\begin{aligned}
&A = \{\text{no dark counts}\},\;
B = \{\text{dark count in }D_1\},\;\\
&C = \{\text{dark count in }D_2\},\;
D = \{\text{dark counts in both detectors}\}.
\end{aligned}
$$
Assigning a random bit value to double clicks and again assuming that the ``correct'' detector should be $D_1$, the conditional error probabilities are
\begin{align}
\left.p_{\mathrm{err}}\right|_A &= p_{01}+\frac{1}{2} p_{11}, \\
\left.p_{\mathrm{err}}\right|_B &= \frac{1}{2}\left(p_{01}+p_{11}\right), \\
\left.p_{\mathrm{err}}\right|_C &= p_{00}+p_{01}+\frac{1}{2}\left(p_{10}+p_{11}\right), \\
\left.p_{\mathrm{err}}\right|_D &= \frac{1}{2}.
\end{align}
The resulting error probability for a Fock state $|n\rangle$ in the $X$ basis is
\begin{equation}
\tilde{H}^X_{n, a}
=\left(1-p_{\mathrm{d}}\right)^2 p_{\mathrm{err}}\big|_A
+p_{\mathrm{d}}\left(1-p_{\mathrm{d}}\right)\left(p_{\mathrm{err}}\big|_B+p_{\mathrm{err}}\big|_C\right)
+p_{\mathrm{d}}^2 p_{\mathrm{err}}\big|_D,
\end{equation}
for all $n\in\mathbb{N}$ and intensities $a\in A$.
Similarly, the click probability (yield) for a Fock state with $n$ photons is
\begin{equation}
\tilde{Y}^{X/Z}_{n, a}
=1-\left(1-p_{\mathrm{d}}\right)^2 p_{00},
\end{equation}
for all $n\in\mathbb{N}$ and $a\in A$.
These Fock-state quantities are used as reference points $\tilde{Y}^{X/Z}_{n,a}$ and $\tilde{H}^X_{n,a}$ in the linearisation of the CS constraints.

To obtain the observed parameters $Z_a$, $X_a$ and $E_a$ that enter the linear constraints of (P1)--(P3), we apply the channel model to PRWCPs. This yields
\begin{align}\
\frac{Z_a}{q_Z^2p_a} &= 1-(1-p_d)^2 e^{-\eta a}, \label{eq:channelModelZ}\\
\frac{X_a}{q_X^2p_a} &= 1-(1-p_d)^2 e^{-\eta a}, \\
\frac{E_a}{q_X^2p_a} &= \frac{p_d^2}{2}
+p_d(1-p_d)\bigl(1+h_{\eta,a,\delta_A}\bigr)
+(1-p_d)^2\left(\frac{1}{2}+h_{\eta,a,\delta_A}-\frac{e^{-\eta a}}{2}\right),
\end{align}
with
\begin{equation}\label{eq:channelModelh}
h_{\eta,a,\delta_A} 
= \frac{1}{2}\left(\exp(-\eta a \cos^2\delta_A)-\exp(-\eta a \sin^2\delta_A)\right).
\end{equation}

\subsection{Fine-grained model-independent variant}

The coarse-grained description \cite{Zapatero2021} above assumes that the conditional photon-number distribution in round $k$ depends only on the current setting $a_k$ through the interval $[a_k^-,a_k^+]$. A more fine-grained model explicitly allows different patterns of past intensity settings to influence the current intensity in different ways. This leads to the following modifications:

\begin{itemize}
    \item The yields and error rates now depend on the history vector $\vec{c}$ of the last $\xi$ settings. We write $Y^{Z/X}_{n,a,\vec{c}}$ and $H^X_{n,a,\vec{c}}$, with $\vec{c}\in A^\xi$.
    \item The deviation parameter $\delta_\text{max}$, and hence $a_k^\pm$, can also depend on $\vec{c}$. 
    \item Observed statistics are also conditioned on the past settings, e.g.\ $Z_{a_k,a_{k-1},\dots,a_{k-\xi}}$. 
    \item A more precise lower bound on the overlap parameter $\tau^\xi_{ab\vec{c},n}$ can be derived \cite{Sixto2022a}, see Eq.~\eqref{eq:generalTau}.
    \item The objective is now the single-photon yield conditioned on the signal intensity and averaged over the past intensity settings. In particular, the objective is directly the quantity that enters the key rate.
\end{itemize}

For $\xi=1$, the fine-grained analogue of (P1) becomes
\begin{equation}
\begin{aligned}
&\min\; q_Z^2p_\mu\sum_{\vec{c}\in A^\xi}p_{\vec{c}} \, p_L(n=1|\mu,\vec{c})\,Y^Z_{1,\mu,\vec{c}} \quad(=Z_{1,\mu})\\
&\text{subject to:}\\[1mm]
&\frac{Z_{a,\vec{c}}}{q_{\mathrm{Z}}^2 p_a p_{\vec{c}}}\geq 
e^{-a^+}Y^Z_{0,a,\vec{c}} + \sum_{n=1}^{N_\text{cut}} \frac{e^{-a^-} (a^{-})^n}{n!} Y^Z_{n,a,\vec{c}},
\quad \forall a\in A,\ \vec{c}\in A^\xi,\\[1mm]
&\frac{Z_{a,\vec{c}}}{q_{\mathrm{Z}}^2 p_a p_{\vec{c}}}\leq 
e^{-a^-}Y^Z_{0,a,\vec{c}}
+\sum_{n=1}^{N_\text{cut}} \frac{e^{-a^+} (a^{+})^n}{n!} Y^Z_{n,a,\vec{c}}
+\left(1-e^{-a^+}-\sum_{n=1}^{N_\text{cut}}\frac{e^{-a^+} (a^{+})^n}{n!} \right),
\quad \forall a,\vec{c},\\[1mm]
& G_{-}\left(Y^Z_{n, a,\vec{c}}, \tau_{a b\vec{c}, n}^{\xi}\right) 
\leq Y^Z_{n, b,\vec{c}} 
\leq G_{+}\left(Y^Z_{n, a,\vec{c}}, \tau_{a b \vec{c}, n}^{\xi}\right),\\
&\hspace{5cm}\forall a,b\in A,\ \vec{c}\in A^\xi,\ a\neq b,\ n=0,\dots,N_\text{cut}, \\[1mm]
&0\leq Y^Z_{n,a,\vec{c}}\leq 1, \quad \forall a\in A,\ \vec{c}\in A^\xi,\ n = 0,\dots,N_\text{cut},
\end{aligned}
\end{equation}
where $p_{\vec{c}}$ is the probability of the history $\vec{c}$, obtained from the i.i.d.\ choice of intensities, $p_{\vec{c}}=\prod_{j}p_{c_j}$.
Replacing $Z$ by $X$ yields the fine-grained analogue of (P2), and replacing $Z_{a,\vec{c}}$ by $E_{a,\vec{c}}$ and $Y^Z_{n,a,\vec{c}}$ by $H^X_{n,a,\vec{c}}$, together with changing $\min$ to $\max$, gives the fine-grained analogue of (P3).

%The overlap parameter is now bounded  by \cite{Sixto2022a}
%\begin{equation}\label{eq:fineGrainedTau}
%\tau_{a b \vec{c},n}^{\xi}
%=\left[1-\sum_{a_i \in A} p_{a_i}\left(e^{-a_i^{-}}-e^{-a_i^{+}}\right)\right]^{2\xi}.
%\end{equation}

%Compared to the coarse-grained expression~Eq.\eqref{eq:coarseGrainedTau}, the prefactors depending on $a$ and $b$ disappear, and $\tau_{ab\vec{c},n}^\xi$ no longer depends on $n$. In our simulations we again keep a common $\delta_\text{max}$, but if $\delta_\text{max}$ depended on the history, the expression would become slightly more involved (see Eq.~\ref{eq:generalTau} below).

%As before, linearisation points for the CS constraints are taken from the channel model, and the same simulation parameters are used as in the coarse-grained case, except for updated values of $\delta_\text{max}$ and $\xi$ where explicitly stated.

\subsection{Truncated-Gaussian correlation model}\label{sec:truncatedGaussianModel}

The previous formulations are model-independent in the sense that they do not assume any specific functional form for $g_{\vec{a}_k}(\alpha_k)$. 
In practice, however, experimental characterisations often suggest that the actual intensity is approximately Gaussian distributed around some mean, with truncation determined by the hardware \cite{Yoshino2018,Trefilov2025}. 
Motivated by this, in \cite{Sixto2022a} also a truncated-Gaussian correlation model is considered.
In this model $g_{\vec{a}_k}(\alpha_k)$ is taken to be a truncated Gaussian with lower and upper truncation points. This function is fully represented by mean intensity, standard deviation and truncation points, which one can measure experimentally using techniques presented in \cite{Trefilov2025}.

In this model, the function $P_n(\cdot)$ in Eq.~\eqref{eq:assumption2-2} is computed by numerically integrating the Poisson distribution over a truncated Gaussian probability density function $g_{\vec{a}_k}$. If the integrals could be evaluated exactly, one could in principle use the exact photon-number probabilities in the linear constraints and dispense with the lower and upper bounds $p_L$ and $p_U$. In practice we keep the bounding approach to account for numerical integration error.

The CS-overlap parameter is then generalised as follows. Let
\[
\vec{v} = (a_{k-\xi}, \dots, a_k),\qquad
\vec{w} = (b_{k-\xi}, \dots, b_k)
\]
be the past intensity patterns up to round $k$, and let
\[
\vec{f} = (a_{k+1}, \dots, a_{k+\xi})
\]
be the future intensity choices, assumed i.i.d.\ with known probabilities $p(a)$. Then
\begin{equation}
\begin{aligned}
\tau_\xi(\vec{v}, \vec{w}) 
=&\Biggl( 
\sum_{\vec{f} \in A^\xi}
\Biggl[
\prod_{j=1}^{\xi}
\Biggl(
\sum_{n=0}^{N} \sqrt{ 
P_n\bigl(a_{k-\xi+j}, \dots, a_k, a_{k+1}, \dots, a_{k+j} \bigr)\,
P_n\bigl(b_{k-\xi+j}, \dots, b_k, a_{k+1}, \dots, a_{k+j} \bigr)
}
\Biggr)
\Biggr]\\
&\hspace{1cm}\cdot
\prod_{j=1}^\xi p(a_{k+j})
\Biggr)^2.
\end{aligned}\label{eq:generalTau}
\end{equation}
Here $P_n(\cdot)$ returns the photon-number probability of emitting $n$ photons given the full conditioning history (this includes averaging over future intensity choices when relevant for the correlation structure). The product over $j$ ranges over future rounds $k+1$ to $k+\xi$, and for each of these rounds, the inner sum over $n$ computes the overlap of the corresponding photon-number distributions. The sum over $\vec{f}\in A^\xi$ averages over all possible future intensity sequences, weighted by their probabilities.

This expression is fully general, and, when lower bounds are substituted for exact ones on the photon-number probabilities, the coarse-grained $\tau$ used in the model-independent case is recovered. As in the model-independent analysis, we typically restrict ourselves to CS constraints between histories $\vec{v}$ and $\vec{w}$ that differ only in the current intensity setting (i.e.\ $a_{k-\xi}=b_{k-\xi},\dots,a_{k-1}=b_{k-1},a_k\neq b_k$), see also \cite{Sixto2022a}. The bounds in the linear constraints are still taken from the channel model described above (see eqs.~(\ref{eq:channelModelZ}-\ref{eq:channelModelh})), where the history dependent intensity $a$ in these equations is set to the mean of  the truncating Gaussian describing $g_{\vec{a}_k}(\alpha_k)$.

\section{Simulations}\label{sec:simulations}

\subsection{Role of IPOPT and outlook.}
The coarse-grained, fine-grained, and truncated-Gaussian formulations above all lead to optimisation problems in which the objective and most constraints are linear, but the CS constraints introduce non-convex square-root nonlinearities. 

In this work we use the interior-point smooth non-linear solver IPOPT \cite{waechter2006} to solve the full nonlinear programs (P1)--(P3) and their fine-grained and truncated-Gaussian variants. These results are then used as reference points for the outer optimisation to obtain the lower bound on the key rate. We note that one can model the non-linearity in Eq. \eqref{eq:sqrt} via the second-order cone, and the region splitting in Eq. \eqref{eq:split} via integer programming, and therefore the entire optimization is naturally formulated as a mixed-integer second-order cone program. Such problems can be solved, for example, by Gurobi \cite{gurobi} or Pajarito \cite{coey2020}, and in principle come with certified lower bounds. However, our numerical experiments showed that this method was slower and less reliable than using IPOPT, and therefore we adopted the latter method for this manuscript.

We compare the resulting key rates to those obtained by the canonical channel-model-based approach of Zapatero \emph{et al.} \cite{Zapatero2021}. 
This allows us to quantify the effect of the CS linearisation on the security analysis and to assess to what extent modern nonlinear optimisation tools can be used to obtain tighter and more reliable bounds in decoy-state QKD with intensity-correlated sources.

\subsection{Coarse grained intensity correlations.}

In this subsection we treat the intensity correlations as in section \ref{subsec:coarseGrainedFormulation} we use a course grained model independent calculations.
We show that our method can outperform the canonical solution and finds an essentially optimal solution.
The following values are used in the simulations to compare the IPOPT-based nonlinear optimisation based choice of linearization points with the canonical points introduced by Zapatero \emph{et al.}:
\begin{itemize}
    \item Intensity choices: $\omega = 10^{-4}$ (vacuum-like), $\nu = 0.1$, $\mu = 0.48$.
    \item Intensity probabilities: $p_\mu\approx 1$, $p_\nu = p_\omega = (1-p_\mu)/2$.
    \item Basis probabilities: $q_Z \approx 1$, $q_X=1-q_Z$.
    \item Detector efficiency: $\eta_\text{det} = 0.65$.
    \item Dark-count probability: $p_d = 7.2\times 10^{-8}$.
    \item Channel transmittance: $\eta_\text{ch}=10^{-0.2 L/10}$, where $L$ is the distance in km.
    \item Misalignment angle: $\delta_A = 0.08$.
    \item Error-correction efficiency: $f_{EC}=1.16$.
    \item Correlation range: $\xi \in \{1,3\}$.
    \item Maximum relative intensity deviation: $\delta_{\max} \in \{10^{-4},10^{-2}\}$.
    \item Photon-number cut-off: $N_\text{cut} = 10$.
\end{itemize}

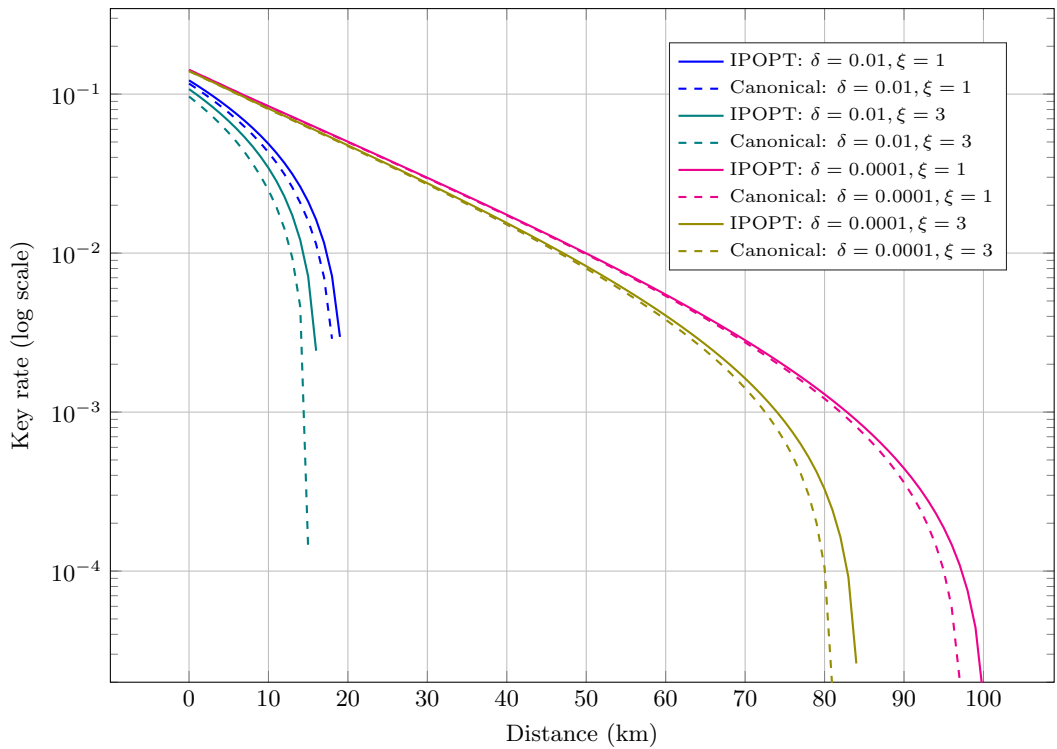
\begin{figure}[ht]
	\centering
 	\begin{tikzpicture}
		\begin{axis}[%
			scale only axis,
			width = 0.7\textwidth,
    		height = 0.5\textwidth,
            ymode = log,
			ymin=2e-5,
			%ymax=1e-5,
			grid=major,
			xlabel = {Distance (km)},
            ylabel = {Key rate (log scale)},
			axis background/.style={fill=white},
			legend style={at={(0.95,0.95)},legend cell align=left, align=left, draw=white!15!black, font=\scriptsize}
			]
            \addplot[color=blue, style=thick] table[col sep=space] {plot_data/comp_i_d01xi1.txt};
            \addlegendentry{IPOPT: $\delta_{\max} = 0.01, \xi = 1$}
            \addplot[color=blue, dashed, style=thick] table[col sep=space] {plot_data/comp_c_d01xi1.txt};
            \addlegendentry{Canonical:  $\delta_{\max} = 0.01, \xi = 1$}
            \addplot[color=teal, style=thick] table[col sep=space] {plot_data/comp_i_d01xi3.txt};
            \addlegendentry{IPOPT: $\delta_{\max} = 0.01, \xi = 3$}
            \addplot[color=teal, dashed, style=thick] table[col sep=space] {plot_data/comp_c_d01xi3.txt};
            \addlegendentry{Canonical: $\delta_{\max} = 0.01, \xi = 3$}
            \addplot[color=magenta, style=thick] table[col sep=space] {plot_data/comp_i_d0001xi1.txt};
            \addlegendentry{IPOPT: $\delta_{\max} = 0.0001, \xi = 1$}
            \addplot[color=magenta, dashed, style=thick] table[col sep=space] {plot_data/comp_c_d0001xi1.txt};
            \addlegendentry{Canonical: $\delta_{\max} = 0.0001, \xi = 1$}
            \addplot[color=olive, style=thick] table[col sep=space] {plot_data/comp_i_d0001xi3.txt};
            \addlegendentry{IPOPT: $\delta_{\max} = 0.0001, \xi = 3$}
            \addplot[color=olive, dashed, style=thick] table[col sep=space] {plot_data/comp_c_d0001xi3.txt};
            \addlegendentry{Canonical: $\delta_{\max} = 0.0001, \xi = 3$}
        \end{axis}
	\end{tikzpicture}
	\caption{Coarse grained key rate calculated using formulation described in Section \ref{subsec:coarseGrainedFormulation}. It is apparent that choosing IPOPT to calculate the reference points for outer linearization outperforms the canonical reference points derived from the noise model. In addition in this particular instance IPOPT essentially finds the optimal points, which can be verified by comparing IPOPT results to outer optimization results using IPOPT results as reference points.}
    \label{fig:canonical vs IPOPT}
\end{figure}

Note that in \cite{Zapatero2021} the intensity choices were optimised for each link distance. Here we keep them fixed for easy reproducibility of our plots. The results are presented in FIG.~\ref{fig:canonical vs IPOPT}. {}{Using our method improves the attainable link length up to $3$~km. Across the compared curves, IPOPT gives an overall key-rate advantage of about $4.36\%$ by area-under-curve, with a typical point-wise improvement of $2.46\%$ (median). The improvement comes at minimal computational cost, as each point in the plots takes only on the order of seconds to calculate on a standard office laptop.}

\subsection{Fine-grained intensity correlations with Gaussian correlation model.}

\begin{figure}
	\centering
 	\begin{tikzpicture}
		\begin{axis}[%
			scale only axis,
			width = 0.7\textwidth,
    		height = 0.5\textwidth,
            ymode = log,
			ymin=1e-7,
			%ymax=1e-5,
			grid=major,
			xlabel = {Distance (km)},
            ylabel = {Key rate (log scale)},
			axis background/.style={fill=white},
			legend style={at={(0.95,0.95)},legend cell align=left, align=left, draw=white!15!black, font=\scriptsize}
			]
            \addplot[color=blue, style=thick] table[col sep=space] {plot_data/trefilov_i43.txt};
            \addlegendentry{IPOPT: $\mu_1 = 0.43, \xi = 1$}
            \addplot[color=blue, dashed, style=thick] table[col sep=space] {plot_data/trefilov_c43.txt};
            \addlegendentry{Canonical: $\mu_1 = 0.43, \xi = 1$}
            \addplot[color=teal, style=thick] table[col sep=space] {plot_data/trefilov_i22.txt};
            \addlegendentry{IPOPT: $\mu_1 = 0.22, \xi = 1$}
            \addplot[color=teal, dashed, style=thick] table[col sep=space] {plot_data/trefilov_c22.txt};
            \addlegendentry{Canonical: $\mu_1 = 0.22, \xi = 1$}
            \addplot[color=magenta, style=thick] table[col sep=space] {plot_data/trefilov_i11.txt};
            \addlegendentry{IPOPT: $\mu_1 = 0.11, \xi = 1$}
            \addplot[color=magenta, dashed, style=thick] table[col sep=space] {plot_data/trefilov_c11.txt};
            \addlegendentry{Canonical: $\mu_1 = 0.11, \xi = 1$}
        \end{axis}
	\end{tikzpicture}
	\caption{Here we compare lower bounds on key rates for System B from \cite{Trefilov2025} obtained by using canonical linearization points and IPOPT. 
    In \cite{Trefilov2025} the Gaussian correlation model could be modified by adding additional global attenuation to control the intensity of the signal state $\mu_1$ while keeping the intensity ratios constant. 
    We illustrate the power of our new method by considering
    system B from \cite{Trefilov2025} with three choices of global attenuation $\mu_1\in\{0.43,0.22,0.11\}$. In all three cases our method significantly outperformed the canonical choice of the reference points.} 
   
    \label{fig:Trefilov}
\end{figure}
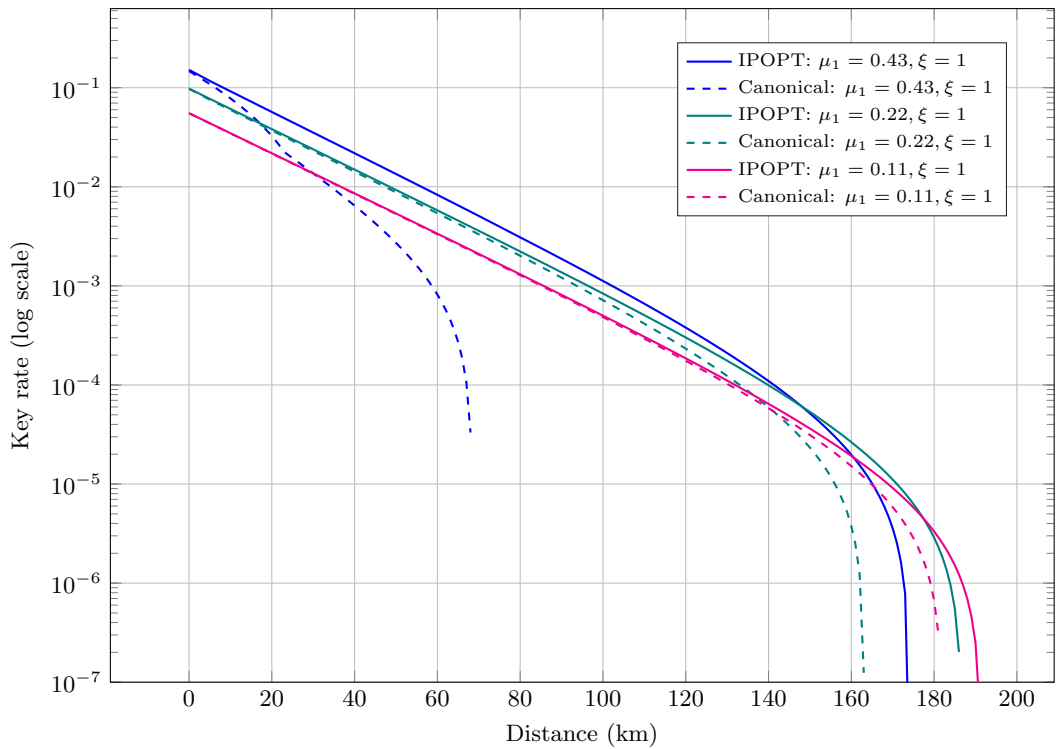

In \cite{Trefilov2025} particular hardware implementations were tested for intensity correlations and the methods described here based on \cite{Sixto2022a} were used to assess the quality of these hardware implementations. A Gaussian correlation model with fine-grained statistics was used.
In this case using the canonical reference points produces much looser key rate bounds as the observed data does not fit to the channel model perfectly. 
Our method can provide superior key rate bounds, especially for particular global attenuation. The results are presented in FIG.~\ref{fig:Trefilov}. In the simulation we use the Gaussian correlation data of ``System B'' from \cite{Trefilov2025} with different values of global attenuation and other system parameters as defined above.

Note that authors of \cite{Trefilov2025} do not use the canonical reference points in their simulations; therefore, FIG.~\ref{fig:Trefilov} should not be interpreted as a direct comparison with simulations presented therein. In \cite{Trefilov2025} the authors present a heuristic to obtain ``good'' reference points in an iterative manner. They start with canonical reference points and then solution obtained in $i^\text{th}$ outside linearization optimization is used as reference points for $i+1^\text{st}$ iteration. Such a method is certainly correct; however, it is much more computationally intensive and less reproducible than our method.

\section{Code availability}

The results of this paper can be reproduced with code available at the following repository: \url{https://github.com/araujoms/qkd_intensity_correlation}.

\section{Conclusion}

We studied decoy-state BB84 implemented with phase-randomized weak coherent pulses whose actual intensities exhibit finite-range correlations across rounds. In this regime the standard decoy-state assumption—intensity-independent yields and error rates—breaks down because intensity information can be indirectly leaked through correlated fluctuations, and the parameter-estimation task must incorporate additional constraints linking different intensity settings. Building on the Cauchy–Schwarz (CS) constraints of Zapatero \emph{et al.} \cite{Zapatero2021}, we formulated the resulting estimation problems as nonlinear programs in which only the CS constraints introduce (square-root) nonlinearities.

Our main contribution is a practical and reproducible way to choose linearisation points for the outer linearisation approach: instead of relying on channel-model-derived “canonical” reference points, we first solve the original problems with the full, non-linearised CS constraints using the interior-point solver IPOPT, and then use the obtained candidate solution as the reference point for the subsequent outer linearisation that certifies security. This two-stage workflow preserves the composable soundness of the linearised outer optimisation, while substantially improving tightness whenever the canonical reference points are poorly matched to the observed data.

Numerical simulations demonstrate both aspects. For the coarse-grained, model-independent formulation we observe consistent improvements over the canonical method and, in the tested regime, near-perfect agreement between the IPOPT solution and the subsequent linearised outer optimisation—providing an explicit certificate that the nonlinear solve reached the true optimum. For the fine-grained truncated-Gaussian correlation model inspired by experimental characterisation presented in \cite{Trefilov2025}, we find even larger gains: when the measured statistics deviate from the simple channel model used to generate canonical points, IPOPT-derived reference points yield markedly tighter lower bounds on the asymptotic key rate, without the need for iterative heuristics.

These results suggest that modern nonlinear optimisation tools can play a direct role in security analyses of practical QKD systems with memory and drift, not by replacing conservative outer-bounding steps, but by supplying high-quality, instance-specific linearisation points that are robust to model mismatch. 
Overall, IPOPT-assisted linearisation provides a simple drop-in upgrade to existing CS-based decoy-state analyses, improving achievable rates while retaining provable security guarantees. 

\section{Acknowledgments}
The research of M.A.~was supported by the Spanish Agencia Estatal de Investigación, Grant Nos.~RYC2023-044074-I and PID2024-161725OA-I00, by the Q-CAYLE project, funded by the European Union-Next Generation UE/MICIU/Plan de Recuperación, Transformación y Resiliencia/Junta de Castilla y León (PRTRC17.11), and also by the Department of Education of the Junta de Castilla y León and FEDER Funds (Reference: CLU-2023-1-05).

M.P.~acknowledges funding by the Austrian Research Promotion Agency (FFG) through the Project NSPT-QKD~FO999915265.

\end{document}